# Electronic transport properties of Ir-decorated graphene

Yilin Wang,[1,2,4] Shudong Xiao,[2] Xinghan Cai,[2] Wenzhong Bao,[2] Janice Reutt-Robey[1,4] and Michael S. Fuhrer[1,2,3*]

[1]*Materials Research Science and Engineering Center, Department of Physics, University of Maryland, College Park, MD 20742, USA*

[2]*Center for Nanophysics and Advanced Materials, University of Maryland, College Park, MD 20742, USA*

[3]*School of Physics, Monash University, Victoria 3800, Australia*

[4]*Department of Chemistry and Biochemistry, University of Maryland, College Park, MD 20742, USA*

* michael.fuhrer@monash.edu

Graphene decorated with 5*d* transitional metal atoms is predicted to exhibit many intriguing properties; for example iridium adatoms are proposed to induce a substantial topological gap in graphene. We extensively investigated the conductivity of single-layer graphene decorated with iridium deposited in ultra-high vacuum at low temperature (7 K) as a function of Ir concentration, carrier density, temperature, and annealing conditions. Our results are consistent with the formation of Ir clusters of ~100 atoms at low temperature, with each cluster donating a single electronic charge to graphene. Annealing graphene increases the cluster size, reducing the doping and increasing the mobility. We do not observe any sign of an energy gap induced by spin-orbit coupling, possibly due to the clustering of Ir.

Graphene, a two-dimensional honeycomb structure of carbon atoms, has been intensively studied due to its novel electronic and structural properties.[1] A striking aspect of graphene is that every atom is a surface atom, and the two-dimensional electron gas in graphene is exposed at the surface. This allows graphene's electronic properties to be tuned by the appropriate introduction of disorder/impurities, such as vacancies,[2] adatoms[3,4,5,6,7] and various molecules.[8,9,10] For example, physisorbed potassium donates an electron to graphene and the ions act as charged impurity scattering centers, decreasing the mobility and conductivity of graphene;[3] while chemisorbed hydrogen and fluorine introduce resonant scattering centers, inducing a band gap and insulating behavior.[11,12] Recently, transition metal adatoms on graphene are of particular interest and have attracted great attention due to a number of fascinating theoretical predictions.[13,14,15] Several 5$d$ metal atoms are expected to induce the quantum spin Hall effect[13] or quantum anomalous Hall effect[14] in graphene due to the enhanced spin-orbit coupling in graphene. Graphene decorated with iridium (Ir) or osmium adatoms is predicted to realize a two-dimensional topological insulator protected by a substantial band gap (~300 meV).[15] These predictions motivate the experimental study of the properties of graphene decorated with 5$d$ heavy metals.[16,17,18,19]

In this work, we investigate the *in-situ* transport properties of single-layer graphene decorated with Ir deposited at low temperature (7 K) under ultra-high vacuum (UHV) conditions. We measure the conductivity as a function of Ir concentration, carrier density, temperature, and annealing conditions. The results are consistent with the formation of clusters of Ir on graphene, even for deposition at low temperature, with each cluster containing ~100 Ir atoms and donating ~1 electron to graphene, acting as a charged impurity scattering center. Annealing Ir-decorated graphene to room temperature greatly reduces the doping and increases the mobility, consistent with greatly increased cluster size. No signature of any significant bandgap in graphene decorated with Ir adatoms was observed and is attribute to the formation of Ir clusters.

**Results**

**Conductivity of as-fabricated and Ir-decorated graphene.** Ir was deposited via electron-beam evaporator in UHV. To vary the coverage, the device was exposed to a controlled flux with sequential exposures at a fixed sample temperature of 7 K. Following each deposition, the conductivity as a function of gate voltage $\sigma(V_g)$ was measured. Figure 1(a) shows $\sigma(V_g)$ for the pristine device and the device with four different Ir doping concentrations. With increasing Ir deposition, several features become apparent: 1) the gate voltage of minimum conductivity $V_{g,\,min}$ shifts to more negative value, 2) the mobility $\mu$ decreases, 3) the minimum conductivity $\sigma_{min}$ decreases. All of these features are similar to the effect of charged impurities on graphene, observed previously by deposition of potassium;[3] we will discuss each in detail below.

**Discussion**

We fit $\sigma(V_g)$ at high $|V_g|$ to

$$\sigma(V_g) = ne\mu = \mu c_g (V_g - V_{g,\min}) + \sigma_{res} \quad (1)$$

separately for electron conduction ($V_g - V_{g,\min} > 0$) and hole conduction ($V_g - V_{g,\min} < 0$) in order to determine the electron and hole mobilities $\mu_e$ and $\mu_h$, the threshold shift $\Delta V_{g,\min}$, and the residual conductivity $\sigma_{res}$, where $n$ is the carrier density, $e$ is the electronic charge and $c_g$ is the gate capacitance per area. Figure 1(b) shows inverse of electron mobility $1/\mu_e$ and hole mobility $1/\mu_h$ versus Ir coverage, both of which are linear, demonstrating the mobility depends inversely on the density of impurities $1/\mu \propto n_{imp}$ (Matthiessen's rule)[20]:

$$\sigma(n) = Ce \left| \frac{n}{n_{imp}} \right| \quad (2)$$

where $C$ is a constant. Although the $\mu_e$ and $\mu_h$ are distinct, their ratio $\mu_e/\mu_h$ remains approximately 0.8 before and with increasing Ir coverage, as shown in the inset of Fig. 1(b). The similar electron-hole asymmetry in mobility is also observed for scattering by potassium adsorbates[3] and follows from the electrostatic environment of the graphene sample.[21] The constant $C$ is 7 x 10$^{17}$ V$^{-1}$s$^{-1}$ (9 x 10$^{17}$ V$^{-1}$s$^{-1}$) for electrons (holes), about two orders of magnitude larger than found for K adatoms, indicating Ir is about 2 order of magnitude less effective at scattering electrons in graphene. Figure 1(c) and 1(d) show $\Delta V_{g,\min}$ and $\sigma_{min}$ as a function of $1/\mu_e$, respectively. The results of both $\Delta V_{g,\min}$ and $\sigma_{min}$ agree well with that of potassium adatoms and can be well described by the previous theoretical predictions generated for impurity charge $Ze$ with $Z = 1$, and impurity–graphene distance $d$ = 0.3 nm - 1.0 nm.[3, 20] Notably, the theoretical predictions are very different for $Z \neq 1$. For example, for a fixed charge transfer $\Delta V_{g,min} = Zn_{imp}$, the scattering cross-section of an impurity scales as $Z^2$ however the density of impurities $n_{imp}$ scales as $1/Z$, hence the mobility scales as $Z$. Thus the results strongly suggest scattering by charged impurities with $Z \approx 1$. Together with the observation of $C$ about two orders of magnitude larger for Ir adatoms than for K adatoms, we infer that scattering is due to clusters[22] of around 100 Ir adatoms with a total charge of ~1 $e$. This is entirely consistent with the observation of $\Delta V_{g,min}$ about two orders of magnitude lower at a given Ir concentration than for a similar concentration of K adataoms. Note also that the charge transferred by Ir in clusters is much smaller than the value for isolated Ir calculated by density functional theory ($Z = 0.22$).[23] It is somewhat surprising that Ir forms clusters of this size so readily on a graphene substrate of $T = 7$ K. However the calculated barrier for an Ir adatom to diffuse through the bridge site on graphene is very small, ~50 meV,[15] and the Ir-Ir binding is stronger than the Ir-C binding,[22] therefore Ir adatoms are highly mobile on graphene and susceptible to form three dimensional clusters even at low temperature.

We also explored very high Ir coverages (>1 ML). Figure 2 shows the shift of $V_{g, min}$ as a function of Ir coverage; $\sigma(V_g)$ was measured during the continuous deposition of Ir. At the beginning of deposition, Ir tends to form uniform clusters randomly distributed on the graphene surface, and $V_{g, min}$ drops fast and is roughly linear with increasing Ir coverage. At higher coverages $V_{g, min}$ drops at a slower rate, consistent with the formation of larger clusters, and finally reaches a saturated value at the Ir coverage of 1.2 ML. This presumably marks the transition from clusters to a continuous film. With increased coverage beyond this point, $V_{g,\,min}$ gradually recovers. The results are similar to those obtained for Pt,[5] where it was also observed that small clusters produced *n*-type doping, with a reduction or even reversal of doping as a continuous film is formed. Monolayer graphene on single crystal Ir is known to be slightly *p*-doped. We conclude that the larger the cluster size, the smaller the charge transfer efficiency. Although $V_{g,\,min}$ goes back above 1.2 ML, the resistivity at Dirac point ($\rho_{xx}$) continues to rise with increasing the Ir coverage, as shown in the inset of Fig. 2. This indicates the failure of charged-impurity scattering to describe the high coverage regime. Presumably other disorder, potentially short ranged scattering, also play an important role in this regime, but this requires additional work to understand.

We also explored temperature as a means to tune the cluster size after deposition.[4] Figure 3(a) shows $V_{g,\,min}$ of graphene with 0.085 ML Ir deposited at $T = 7$ K [$\sigma(V_g)$ data shown in Fig. 1(a)] as a function of temperature. During warming, $V_{g,\,min}$ first shifts slightly towards negative gate voltage and then for $T > 90$ K shifts more rapidly to positive, eventually reaching its initial value before Ir deposition. The $V_{g,\,min}$ shift reflects the rearrangement of Ir clusters, while the diffusion, growth and nucleation of atoms on surface is a complex process, we speculate that below 90 K, Ir clusters do not grow appreciably, while some movement of individual adatoms which leads to the negative gate voltage shift of $V_{g,\,min}$. Above 90 K, Ir clusters grow by Ostwald ripening, reducing the charge transfer efficiency and positively shifting $V_{g,\,min}$. Unsurprisingly, formation of large clusters at higher temperature (350 K) is found to be irreversible when re-cooling to low temperature, i.e. the ripening process is irreversible, as shown in Fig. 3(a). Figure 3(b) shows a comparison of σ(V$_g$) for pristine graphene and Ir-decorated graphene annealed to form large clusters. $V_{g, min}$ remains almost the same, indicating there is no change in charge transfer between large Ir clusters and graphene Consistent with charged-impurity-dominated scattering, $σ_{min}$ is nearly unchanged, and $μ$ barely decreases, from 7000 to 6000 cm$^2$/Vs. In contrast to small clusters, large clusters have low impact on the conductivity of graphene. The results are similar to those seen with Au clusters[4, 7] on graphene.

We further studied the temperature dependence of ρ$_{xx}$ of graphene decorated with Ir to search for an energy gap induced by spin-orbit coupling.[15] As shown in Fig. 4, ρ$_{xx}$ increases with decreasing temperature, and is well described by ρ$_{xx}$(T) ~ ln(T). We also plot ρ$_{xx}$ in logarithmic scale as a function of inverse temperature (inset of Fig. 4) and find ρ$_{xx}$(T) is poorly described by the simple thermal activation model $\rho_{xx} \propto e^{E_g/k_B T}$; The obtained fitting gap is extremely small, $E_g$ < 1 meV, which is nonphysical, in that it is smaller than the measurement temperature $k_B T$ and much smaller than the disorder energy scale of order 50 meV. The roughly logarithmic ρ$_{xx}$(T) may originate from increased weak localization in graphene, as have been observed in other noble metal-decorated graphene.[24] The enhanced spin-orbit coupling in Ir-decorated graphene was also not seen by the non-local transport measurement, as discussed elsewhere.[25] We speculate that the failure to observe the predicted enhanced spin-orbit coupling and substantial energy gap in in Ir-decorated graphene is because of the formation of Ir clusters on graphene, which is different from the single adatom model used in the theory.[15] Adatom clustering has also been shown to have a detrimental effect on the formation of the topological phase since the induced spin-orbit coupling vanishes in the region between the islands, which leads to the failure to observe the topological bulk gap.[26]

In summary, we have investigated the electronic transport properties of graphene decorated with 5$d$ transitional metal Ir. Ir tends to form clusters on graphene, acting as charged impurity scattering centers with a single electronic charge per cluster. No topological gap induced by spin-orbit coupling is observed, either due to the lack of such a gap in graphene with clustered Ir, or the lack of a global gap in transport due to inhomogeneity in graphene with adatom clusters. These findings provide guidance for future experiments aimed at achieving strong spin-orbit coupling in metal-decorated graphene.

**Methods**

**Graphene devices fabrication and Electrical transport measurements.** Graphene flakes are obtained by mechanical exfoliation of graphite on a 300-nm-$SiO_2$/Si substrate and are identified by color contrast in optical microscope imaging and confirmed by Raman spectroscopy. The electrical contacts are defined with standard electron beam lithography and thermally evaporated Cr/Au (5 nm/100 nm). After annealing in $H_2$/Ar gas at 350 °C to remove resist residue,[27] the device was mounted on a cryostat in an UHV chamber. All measurements were taken by using a conventional four-probe lock-in technique with a low frequency of 3.7 Hz.

**Acknowledgments**

This work was supported by the NSF-MRSEC at the University of Maryland, DMR 0520471. M.S.F. is supported by the Australian Research Council.

**Author contributions**

Y.W., M.S.F. and J.R.R. conceived the experiments. Y.W., X.C. and W.B. fabricated graphene devices. Y.W. and S.X. performed the measurements. Y.W., M.S.F. and J.R.R. analyzed the data, discussed the results and wrote the manuscript.

Figure Captions :

**Figure 1 | Conductivity evolution after Ir deposition.** (a) The conductivity σ versus gate voltage $V_g$ for pristine graphene and at four different Ir coverages taken at 7 K. Here 1 ML =1.56 x $10^{15}$ cm$^{-2}$ is defined from the atomic density of the Ir(111) surface. (b) Inverse of electron mobility $1/\mu_e$ and hole mobility $1/\mu_h$ versus Ir coverage. Lines are linear fits to all data points. Inset: The ratio of $\mu_e$ to $\mu_h$ versus Ir coverage. The blue line corresponds to the electron-hole asymmetry observed for potassium in Ref. 3. (c) The shift of gate voltage of minimum conductivity -$\Delta V_{g,min}$ as a function of $1/\mu_e$, which is proportional to the impurity concentration. All -$\Delta V_{g,min}$ values are offset by 2 V to account for initial disorder. (d) The minimum conductivity $\sigma_{min}$ as a function of $1/\mu_e$. Lines in (c) and (d) correspond to the theory in Ref. 3.

**Figure 2 | Ir coverage dependence of the shift of gate voltage of minimum conductivity of graphene $\Delta V_{g,min}$.** Inset: The resistivity of graphene at $V_g = V_{g,min}$ as a function of Ir coverage. The measurement was carried out during the continuous deposition of Ir.

**Figure 3 | Temperature dependence of $V_{g,min}$ and σ($V_g$).** (a) $V_{g,\ min}$ of graphene decorated with 0.085 ML Ir as a function of temperature. (b) A comparison of σ($V_g$) for pristine graphene and for 0.085 ML Ir-decorated graphene deposited at 7 K and annealed at 350 K. Data was taken at 7 K.

**Figure 4 | Temperature dependence of $\rho_{xx}$ at $V_g$ = $V_{g,min}$ for 0.085 ML Ir-decorated graphene.** Inset: Temperature dependence of $\rho_{xx}$ at $V_g = V_{g,min}$ for 0.085

ML Ir-decorated graphene. The red line is a fit to the thermal activation model as described in the text.

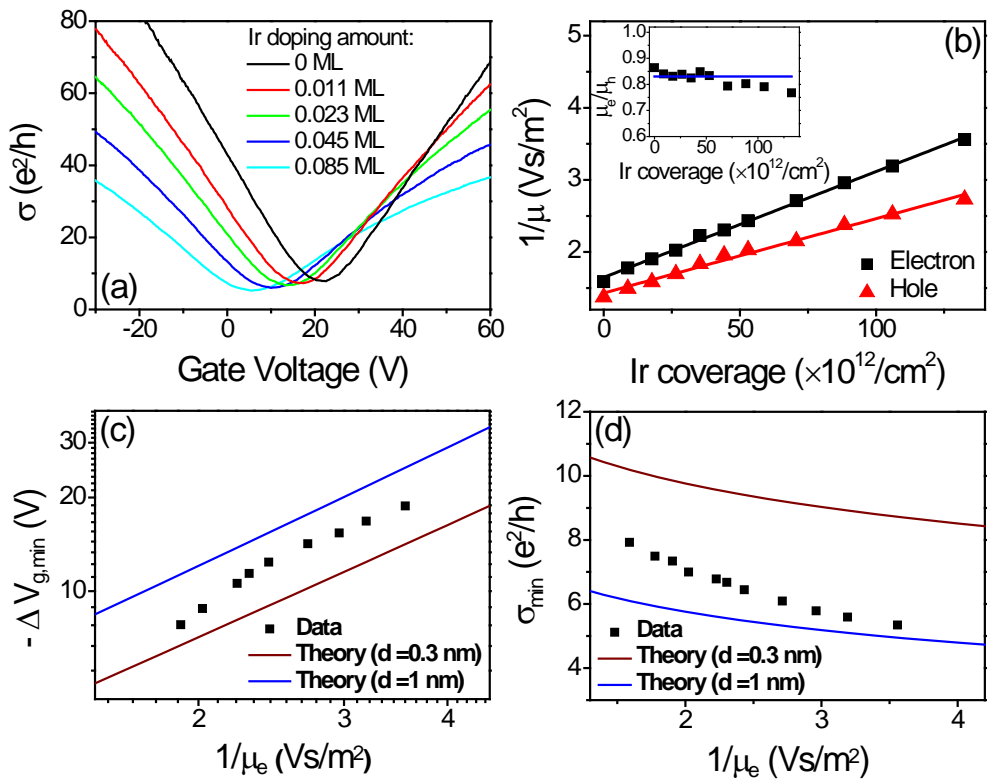

Fig. 1, Y.L. Wang *et al*

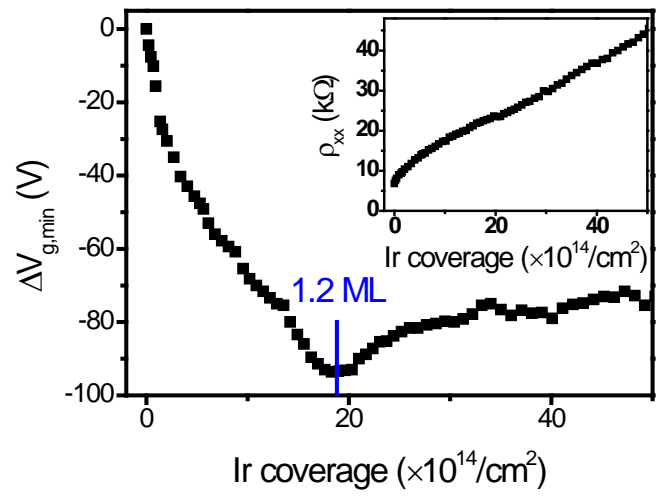

Fig. 2, Y.L. Wang *et al*

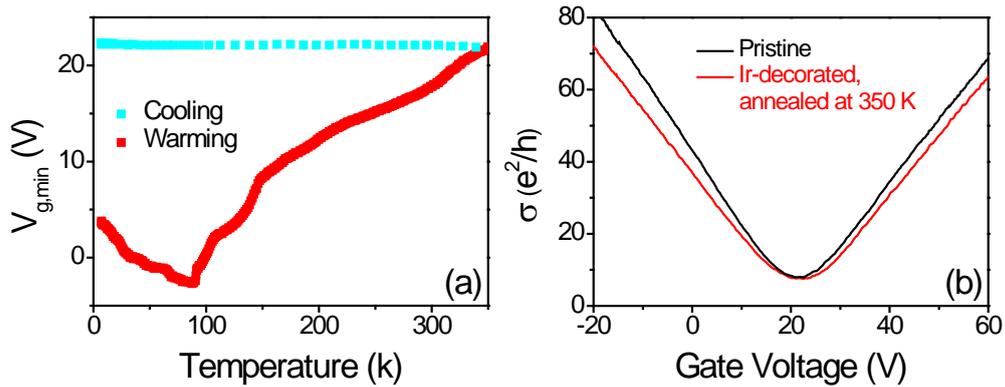

Fig. 3, Y.L. Wang *et al*

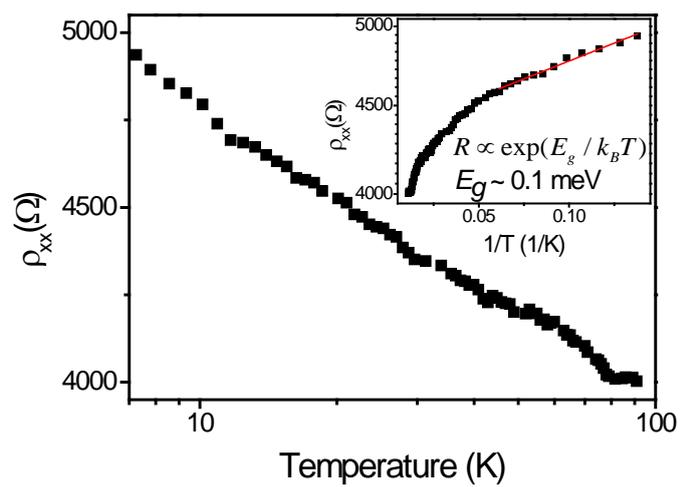

Fig. 4, Y.L. Wang *et al*